\documentclass[aps,pre,reprint,onecolumn,notitlepage,groupedaddress,amsmath]{revtex4-2}

\usepackage{graphicx}% Include figure files
\usepackage{dcolumn}% Align table columns on decimal point
\usepackage{bm}% bold math
\usepackage{hyperref}% add hypertext capabilities
\usepackage{mathrsfs}
\usepackage{subfig}
\usepackage{color}
\usepackage{sansmath}
\usepackage{multirow}

\begin{document}

%\preprint{APS/123-QED}

\title{On the hydrodynamics of active particles in viscosity gradients}

\author{Vaseem A. Shaik}
\author{Gwynn J. Elfring}%
 \email{gelfring@mech.ubc.ca}
\affiliation{Department of Mechanical Engineering, Institute of Applied Mathematics,\\
University of British Columbia, Vancouver, BC, V6T 1Z4, Canada}

\date{\today}

\begin{abstract}
In this work, we analyze the motion of an active particle, modeled as a spherical squirmer, in linearly varying viscosity fields. In general, the presence of a particle will disturb a background viscosity field and the disturbance generated depends on the boundary conditions imposed by the particle on the viscosity field. We find that, irrespective of the details of the disturbance, active squirmer-type particle tend to align down viscosity gradients (negative viscotaxis). However, the rate of rotation and the swimming speed along the gradient do depend on the details of the interaction of the particle and the background viscosity field. In addition, we explore the relative importance on the dynamics of the local viscosity changes on the surface of active particles versus the (nonlocal) changes in the flow field due to spatially varying viscosity (from that of a homogeneous fluid). We show that the relative importance of local versus nonlocal effects depends crucially on the boundary conditions imposed by the particle on the field. This work demonstrates the dangers in neglecting the disturbance of the background viscosity caused by the particle as well as in using the local effects alone to capture the particle motion.
\end{abstract}

\maketitle

%\tableofcontents
%\newpage

\section{Introduction}
Active particles are particles or organisms that convert the stored energy or energy from the surroundings into directed motion \citep{Schweitzer2007}, and a suspension of active particles can be referred to as active matter. We are concerned here with the micron-sized active particles such as biological swimming microorganisms or the synthetic microrobots. Because of the ubiquity of microorganisms in nature, as well as aided by tremendous progress in microfluidic experimental techniques, there has been an enormous amount of research on the motion of active particles in viscous fluids, including developing an understanding of the physics governing microorganism motility \citep{Brennen1977, Berg2004, Fauci2006, Lauga2009, Koch2011, Guasto2012, Elgeti2015, Lauga2016, Lauga2020}.

% or to develop novel biomedical procedures (seems no reference is pertinent to this).

Active particles typically reside in the gradients of heat, light or chemicals and often react to these fields by reorienting and moving along the gradients; a phenomenon known as \textit{taxis}. By exploiting this tendency, one can sort or control the active matter by imposing external gradients. A relatively unexplored form of taxis is viscotaxis---directed motion in viscosity gradients. Several microorganisms move through viscosity gradients and hence exhibit viscotaxis. For instance \textit{Leptospira} and \textit{Spiroplasma} have been observed to display positive viscotaxis, moving up viscosity gradients \citep{Kaiser1975, Petrino1978, Daniels1980, Takabe2017}  while \textit{Escherichia coli} displays negative viscotaxis \citep{Sherman1982}. Green microalgae, \textit{Chlamydomonas reinhardtii}, display complex dynamics in the presence of spatial variations in viscosity. If viscosity gradients are weak, they tend to accumulate in the high viscosity regions due to slower speeds, but in the presence of strong viscosity gradients, they reorient towards the low viscosity regions, displaying negative viscotaxis \citep{Stehnach2020, Coppola2021}. \textit{Helicobacter pylori}, a bacterium commonly found in our guts, also swims through viscosity gradients as it propels by locally lowering the viscosity of the surrounding mucus layer \citep{Montecucco2001, Celli2009}. Hence, a physical description of the fluid dynamics of viscotaxis allows us to understand the consequences on the motility of microorganisms in nature, but also presents a mechanism by which to control active matter, both natural and synthetic.

Several researchers have analyzed the motion of particles, passive or active, in viscosity gradients. Initial research in this field focused on particles moving through a fluid that is otherwise homogeneous in the absence of the particle, but the presence of the particle generates a disturbance (inhomogeneity) in fluid properties such as viscosity. For instance, a particle hotter than the surrounding fluid disturbs the temperature and thus the viscosity, thereby generating the gradients. Research of this sort started with the calculation of the force and torque acting on a hot passive particle \citep{Oppenheimer2016}. It was found that a dipolar temperature distribution on the particle, that causes a similar viscosity distribution in the nearby fluid, induces a coupling between the force (resp. torque) and the angular velocity (resp. translational velocity). Such coupling is absent for non-skew particles (eg: sphere, spheroid, disk etc) in the homogeneous Newtonian fluids \citep{Happel1981}. Later, the motion of an active particle in viscosity gradients caused by variation in the concentration of nearby nutrients was analyzed \citep{Shoele2018}. This work modeled active particles as a spherical squirmer with fixed power and assumed a weak dependence of viscosity on the nutrient concentration. In the squirmer model, an active particle propels due to prescribed slip on its surface, which is ultimately a manifestation of the coordinated beating of cilia on the surface of a ciliated organism or the chemical reactions occurring on the surface of a phoretic particle. It was found that the speed of a swimmer can either increase or decrease depending on the relative importance of the advective to the diffusive transport rate of the viscosity. Recently, the motion of a spheroidal squirmer in the nutrient induced viscosity gradients was also analyzed \citep{Eastham2020}. Accounting for a strong coupling between the nutrient concentration and the viscosity, it was found that nutrient advection and viscosity gradients significantly affect the swimming and feeding performance but not the aspect ratio of the particle.

%the speed of a \textcolor{red}{treadmill swimmer (squirmer with simplest slip velocity)} is always enhanced in the viscosity gradients while 

Recent research explores the motion of active particles in preexisting (or background) viscosity gradients. Such analysis is generally carried out by representing active particles as prototypical model swimmers and often by treating the background viscosity field as linear. Initial research in this field modeled active particles as spheres connected by rods, where each sphere locally sees a constant background viscosity, and is acted on by a fixed active (thrust) force, while hydrodynamic interactions are neglected \citep{Liebchen2018}. Using this approach, it was found that non-chiral (linear) active particles generically move up gradients unless they are uniaxial, in which case they do not exhibit viscotaxis by symmetry. Otherwise, non-linear or chiral swimmers can even move down the gradients. As \textit{Leptospira} and \textit{Spiroplasma} exhibit non-uniaxial shapes while \textit{E. coli} is chiral, this work possibly explains the viscotaxis of these organisms based on their shape and the resulting hydrodynamics. Later, the motion of spherical squirmers in viscosity gradients was analyzed, taking into account locally varying hydrodynamic forces and modifications in the flow due to viscosity differences, and they were found to move down viscosity gradients \citep{Datt2019}. A different model microorganism, Taylor's swimming sheet that propels by passing traveling waves along its surface, moving along or against the viscosity gradients was also analyzed \citep{Dandekar2020}. It was found to speed up in the presence of viscosity gradients irrespective of its direction of propulsion. Recently, the axisymmetric motion of a synthetic helical swimmer crossing a sharp as well as diffuse viscosity interface was analyzed \citep{Lopez2020}. It was found that a head-first (resp. tail first) swimmer moving up the gradients experiences the speed reduction (resp. enhancement) but a swimmer moving down the gradient always undergoes the speed reduction irrespective of its orientation.

A viscosity gradient can arise due to the spatial variation of a scalar like the temperature, salt or nutrient concentration in a fluid that is ultimately coupled to viscosity. However, the introduction of a particle will generally disturb a preexisting viscosity field in order to satisfy boundary conditions for the scalar field on the surface of the particle (because the properties of the particle are different from those of the fluid). For instance, even at steady-state a particle in an otherwise linear temperature field will disturb the background field unless the thermal conductivity of the particle and the fluid are identical. The disturbance generated by the particle tends to vanish far from the particle, however, near the particle, it is the same order of magnitude as that of the preexisting (background) viscosity field. Despite this, most previous work on active particles in viscosity gradients have neglected the disturbance caused by the particle to the ambient viscosity field \citep{Liebchen2018, Datt2019, Stehnach2020, Lopez2020}. In this work we aim to address this gap and investigate the effect of the disturbance of the viscosity induced by the presence of the particle and the consequence on viscotaxis. Furthermore, several previous studies used only the local effects of changes in viscosity not the (non-local) modification of the flow (from that in a homogeneous fluid), in order to study the particle motion in viscosity gradients \citep{Liebchen2018, Lopez2020}. These non-local effects may be as important as the complementary local effects. In fact, the non-local effects have been observed to be more important than the local effects for particles (passive and active) in non-Newtonian fluids \citep{Einarsson2017, Riley2017, Gomez2017, Pietrzyk2019}. In light of this, we perform a quantitative comparison between local and non-local effects for the motion of an active particle in viscosity gradients.

We organize the paper as follows. We provide the mathematical formulation associated with the motion of active particles in viscosity gradients in Sec.~\ref{sec:Prob_Form}, and describe the solution methodology in Sec.~\ref{sec:Sol_Method}. We then discuss the influence of disturbance viscosity in Sec.~\ref{sec:Dist_Visc}, and distinguish between the local and non-local effects, and quantify these effects in Sec.~\ref{sec:LocalNL}. We finally provide few concluding remarks in Sec.~\ref{sec:Conc}.

\section{\label{sec:Prob_Form}Swimming in viscosity gradients}

\begin{figure}
    \centering
    \includegraphics[scale = 0.55]{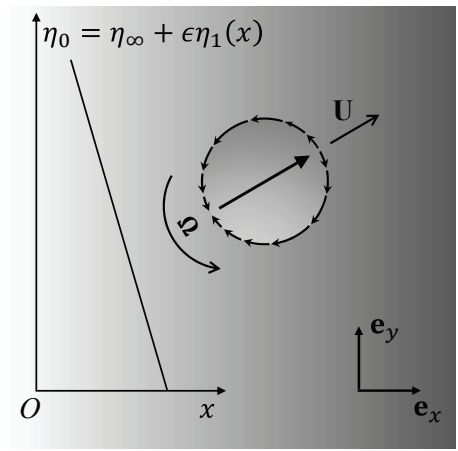}
    \caption{A schematic showing the motion of an active particle in linear viscosity fields and the associated coordinate system. The radius of the particle is $a$. It translates and rotates with the velocities ${\bf{U}}$, ${\bf{\Omega }}$, respectively. The ambient viscosity $\eta_0$ increases linearly with $x$.}
    \label{fig:schematic}
\end{figure}

We consider the motion of an active particle in an otherwise quiescent Newtonian fluid (see Fig.~\ref{fig:schematic} for a schematic). In the absence of the particle, the fluid viscosity, ${\eta _0}\left( {\bf{x}} \right)$, is nonuniform due to corresponding spatial variations in the temperature, salt or nutrient concentration. The particle disturbs the ambient (or background) viscosity field as its properties (like thermal conductivity) are usually different from those of fluid. We denote this disturbance viscosity by $\eta '\left( {\bf{x}} \right)$ and hence the viscosity in the presence of particle is $\eta \left( {\bf{x}} \right) = {\eta _0}\left( {\bf{x}} \right) + \eta '\left( {\bf{x}} \right)$. We are interested in leading order effects and so simply prescribe a linear variation of the ambient viscosity field, ${\eta _0} = {\eta _\infty } + \eta_{\infty} \frac{x}{L}$, where the variation itself occurs on the macroscopic length scale $L$. Usually, this length scale is much larger than the particle size $a$, i.e., $\epsilon  = \frac{a}{L} \ll 1$. This means that in the vicinity of the particle, the ambient viscosity varies weakly, ${\eta _0} = {\eta _\infty } + \frac{{\eta_{\infty} a}}{L}\frac{x}{a} = {\eta _\infty } + \epsilon \eta_{\infty} x/a = {\eta _\infty } + \epsilon {\eta _1}$, where ${\eta _1} = \eta_{\infty} x/a$. Noting the absolute value of the viscosity does not affect the dynamics of an active particle with a prescribed gait \citep{Lauga2009}, we choose an arbitrary point in the fluid as the origin in the lab frame of reference, and we denote the position vector with respect to the origin and the unit vectors in the Cartesian coordinate system, respectively, by ${\bf{x}} = \left( {x,y,z} \right)$, $\left\{ {{{\bf{e}}_x},{{\bf{e}}_y},{{\bf{e}}_z}} \right\}$. In this frame, ${\bf{x}} = {{\bf{x}}_c}$ denotes the center of the particle. Also ${\eta _\infty }$ is some reference viscosity while $\epsilon$ characterizes the deviation from this reference viscosity.

Neglecting fluid inertia, the flow is governed by the incompressible Stokes equations 
\begin{align}
\nabla  \cdot {\boldsymbol{\sigma }} &= {\bm{0}},\\
\nabla  \cdot {\bf{u}} &= 0.
\end{align}
where $\bf{u}$ is the velocity field and $\boldsymbol{\sigma }$ the stress tensor defined as
\begin{equation}
{\bm{\sigma }} =  - p{\bf{I}} + {\eta _\infty }{\bm{\dot \gamma }} + {{\bm{\tau }}_{NN}},
\end{equation}
where ${\bm{\dot \gamma }} = \nabla {\bf{u}} + {\left( {\nabla {\bf{u}}} \right)^T}$ and we have defined
\begin{align}
{{\bm{\tau }}_{NN}} = \left( {\eta \left( {\bf{x}} \right) - {\eta _\infty }} \right){\bm{\dot \gamma }},
\end{align}
which captures the difference in stress due to changes in viscosity from the reference value $\eta_\infty$.

The transport of the scalar (e.g., temperature, salt or nutrient concentration) that determines the viscosity is governed by an advection-diffusion equation. For weak variations of the scalar, the change in the viscosity is linearly proportional to the change in the scalar which means that the viscosity transport is also governed by a similar advection-diffusion equation. The relative order of magnitude of the advective to the diffusive transport of viscosity is characterized by the Péclet number, $Pe$. This number is small for small particles moving at slow speeds in comparison to a highly diffusive scalar like temperature. In this case, the viscosity transport is simply governed by a Laplace equation,
\begin{equation}
{\nabla ^2}\eta  = 0.
\end{equation}
%Similar to the Whitehead's paradox problem for heat transfer \citep{Leal2007}, in the small advection analysis carried out here, howsoever small advection may be near the particle, advection becomes the same order of magnitude as that of diffusion far away from the particle. However, it is not necessary to analyze this far-field region because the contribution of this region to the swimming velocities is $O\left( {\epsilon P{e^m}} \right)$, $m > 0$ and is negligible compared to the first order (in $\epsilon$) swimming velocities found here \citep{Oppenheimer2016}. 

Far away from the particle, the flow decays to zero while the viscosity approaches the ambient viscosity
\begin{equation}
{\bf{u}} \to {\bm{0}},\,\,\eta  \to {\eta _0}\left( {\bf{x}} \right)\,\,{\rm{as}}\,\,r = \left| {\bf{r}} \right| \to \infty,
\end{equation}
where ${\bf{r}} = {\bf{x}} - {{\bf{x}}_c}$. On the particle, the fluid velocity is same as the particle surface velocity as a consequence of the kinematic and no-slip boundary conditions. The particle itself is propelling with a translational velocity ${\bf{U}}$ and an angular velocity ${\bm{\Omega }}$ due to the activity that is embedded in the slip velocity on the particle surface ${{\bf{u}}^s}$. Thus the fluid velocity on the particle is
\begin{equation}
{\bf{u}} = {\bf{U}} + {\boldsymbol{\Omega }} \times {\bf{r}} + {{\bf{u}}^s}\,\,{\rm{on}}\,\,{S_p},
\end{equation}
where $S_p$ denotes the particle surface. We neglect the particle inertia and assume the particle is neutrally buoyant. This implies the hydrodynamic force and torque acting on the particle are zero.
\begin{align}
\bf{F}&=\int_{{S_p}} {{\bf{n}} \cdot {\bm{\sigma }}\,dS}  = {\bm{0}},\\
\bf{L}&=\int_{{S_p}} {{\bf{r}} \times \left( {{\bf{n}} \cdot {\bm{\sigma }}} \right)\,dS} = {\bm{0}},
\end{align}
where ${\bf{n}}$ is an unit normal to the surface that points into the fluid.

We model the active particle as a spherical squirmer of radius $a$ that has only the tangential squirming modes \citep{Lighthill1952, Blake1971, Ishikawa2006}. This model is a good representation of the ciliated organisms like \textit{Paramecium} or \textit{Opalina} which generate slip through a synchronous motion of a large number of cilia on their surface. In this model, the slip velocity ${{\bf{u}}^s}$ is given by
\begin{equation}
{{\bf{u}}^s} =  - \sum\limits_{n = 1}^\infty  {{B_n}{W_n}\left( {{\bf{p}} \cdot {{\bf{e}}_r}} \right){\bf{p}} \cdot \left( {{\bf{I}} - {{\bf{e}}_r}{{\bf{e}}_r}} \right)} ,\quad {W_n}\left( x \right) = \frac{2}{{n\left( {n + 1} \right)}}{P_n}'\left( x \right).
\end{equation}
Here ${\bf{p}}$ is the orientation of the particle, ${{\bf{e}}_r} = {\bf{r}}/r$, ${P_n}$ is the Legendre polynomial of degree $n$ and $B_n$ are the squirming modes. In homogeneous Newtonian fluids, only the $B_1$ mode contributes to the swimming speed while the $B_2$ mode leads to the slowest decaying flow field and hence it determines the far-field representation of the swimmer. The ratio $\alpha = B_2/B_1$ can be used to distinguish the three types of swimmers. So-call pusher swimmers, like \textit{Escherichia coli}, push the fluid along their axis while drawing fluid from their sides. Pullers, like \textit{Chlamydomonas}, do the opposite, in that they pull the fluid along their axis while ejecting fluid from their sides. Neutral swimmers like \textit{Volvox carteri} exhibit the flow signature similar to that of a potential dipole. Puller, pusher and neutral swimmers are characterized by $\alpha > 0$, $< 0$, and $= 0$, respectively.

\section{\label{sec:Sol_Method}Solution Method}

We apply the reciprocal theorem to the unknown flow of the present problem and the auxiliary flow due to a rigid body motion of a spherical particle in a homogeneous fluid of viscosity $\eta_{\infty}$. After making use of the boundary conditions on the particle surface in both problems, we find the swimming velocity of the active particle to be
\begin{equation}
{\bm{\mathsf{U}}} = {{\bm{\mathsf{{{\hat R}}_{FU}^{-1}}}}} \cdot \left[ \int_{{S_p}} {{{\bf{u}}^s} \cdot \left( {{\bf{n}} \cdot {\bm{\mathsf{{\hat T}_U}}}} \right)dS}  + {{\bm{\mathsf{F}}}_{NN}^l} + \int {\left( {\nabla  \cdot {{\bm{\tau }}_{NN}}} \right) \cdot {\bm{\mathsf{{\hat G}_U}}}\,\,dV}\right]. 
\label{eqn:recip_theor}
\end{equation}
Here ${\bm{\mathsf{U}}} = {\left[ {{\bf{U}}\,\,{\bf{\Omega }}} \right]^\top}$ is a six-dimensional vector containing the particle's translational and angular velocity while ${{\bm{\mathsf{F}}}_{NN}^l} = {\left[ {{{\bf{F}}_{NN}^l}\,\,{{\bf{L}}_{NN}^l}} \right]^\top}$ is also a six-dimensional vector consisting of the force and torque due to the stress ${{\bm{\tau }}_{NN}}$
\begin{align}
{{\bf{F}}_{NN}^l} &= \int_{S_p} {\bf{n}}\cdot{{\bm{\tau }}_{NN}}\,dS,\\
{{\bf{L}}_{NN}^l} &= \int_{S_p} {\bf{r}}\times({\bf{n}}\cdot{{\bm{\tau }}_{NN}})\,dS.
\end{align}
We note that the variables with caret are associated with the auxiliary flow problem. Specifically, ${{\bm {\mathsf{{{\hat G}}_{U}}}}} \cdot {\bm{\mathsf{\hat U}}}$, ${{\bm{\mathsf{{{\hat T}}_U}}}} \cdot {\bm{\mathsf{\hat U}}}$, respectively, are the flow field, and the stress tensor due to the rigid body motion of a sphere in a homogeneous fluid. The sphere experiences the hydrodynamic force or torque, ${\bm{\mathsf{\hat F}}} =  - {{\bm{\mathsf{\hat R}}}_{{\bm{\mathsf{FU}}}}} \cdot {\bm{\mathsf{\hat U}}}$. A detailed derivation may be found elsewhere \citep{Elfring2017}. Equation \eqref{eqn:recip_theor} explicitly divides the dynamics of an active particle in fluid with spatially varying viscosity into three contributions: the first term in the brackets on the right-hand side (when multiplied the the mobility) gives the translational and rotational velocity of the active particle in a homogeneous fluid with constant viscosity; the second term represents the effect on the dynamics due to changes in the force and torque from that in a homogeneous fluid; the final term accounts for changes in the dynamics due to changes in the flow field that arise as a result of differences in the stress tensor from that of a homogeneous fluid with constant viscosity.

Separating the above equation \eqref{eqn:recip_theor} for the translational velocity and the angular velocity we have
\begin{equation}
{\bf{U}} = {\bf{\hat R}}_{FU}^{ - 1} \cdot \left[ {\int_{{S_p}} {{{\bf{u}}^s} \cdot \left( {{\bf{n}} \cdot {{{\bf{\hat T}}}_U}} \right)dS}  + \int_{S_p} {\bf{n}}\cdot{{\bm{\tau }}_{NN}}\,dS + \int {\left( {\nabla  \cdot {{\bm{\tau }}_{NN}}} \right) \cdot {{{\bf{\hat G}}}_U}\,dV} } \right],
\end{equation}
\begin{equation}
{\bf{\Omega }} = {\bf{\hat R}}_{L\Omega }^{ - 1} \cdot \left[ {\int_{{S_p}} {{{\bf{u}}^s} \cdot \left( {{\bf{n}} \cdot {{{\bf{\hat T}}}_\Omega }} \right)dS}  + \int_{S_p} {\bf{r}}\times({\bf{n}}\cdot{{\bm{\tau }}_{NN}})\,dS + \int {\left( {\nabla  \cdot {{\bm{\tau }}_{NN}}} \right) \cdot {{{\bf{\hat G}}}_\Omega }\,dV} } \right].
\end{equation}
Here, ${{\bf{\hat G}}_U} \cdot {\bf{\hat U}}$, ${{\bf{\hat T}}_U} \cdot {\bf{\hat U}}$ (resp. ${{\bf{\hat G}}_\Omega } \cdot {\bf{\hat \Omega }}$, ${{\bf{\hat T}}_\Omega } \cdot {\bm{\hat \Omega }}$) are the flow field and the stress tensor due to the translation (resp. rotation) of a sphere in a homogeneous fluid with velocity ${\bf{\hat U}}$ (resp. ${\bf{\hat \Omega }}$). The translating particle experiences the hydrodynamic force ${\bf{\hat F}} =  - {{\bf{\hat R}}_{FU}} \cdot {\bf{\hat U}}$ while the rotating particle experiences the hydrodynamic torque ${\bf{\hat L}} =  - {{\bf{\hat R}}_{L\Omega }} \cdot {\bf{\hat \Omega }}$. For a sphere, ${{\bf{\hat R}}_{FU}} = 6\pi {\eta _\infty }a{\bf{I}}$ and ${{\bf{\hat R}}_{L\Omega }} = 8\pi {\eta _\infty }{a^3}{\bf{I}}$.

We perform a regular perturbation in $\epsilon$ and expand any variable $f(\epsilon)$ as $f = {f_0} + \epsilon {f_1} + {\epsilon ^2}{f_2} + ...$. As $\eta \left( {\bf{x}} \right) - {\eta _\infty } = \epsilon {\eta _1} + \eta '\sim O\left( \epsilon  \right)$, ${{\bm{\tau }}_{NN}} = \left( {\eta \left( {\bf{x}} \right) - {\eta _\infty }} \right){\bm{\dot \gamma }}\sim O\left( \epsilon  \right)$, at leading order, we have a swimmer moving through a homogeneous Newtonian fluid of viscosity $\eta_{\infty}$. The velocity of such swimmer is well known:
\begin{equation}
{{\bf{U}}_0}  = \frac{1}{{6\pi {\eta _\infty }a}}\int_{{S_p}} {{{\bf{u}}^s} \cdot \left( {{\bf{n}} \cdot {{{\bf{\hat T}}}_U}} \right)dS}  = \frac{2}{3}{B_1}{\bf{p}} \equiv {{\bf{U}}_N},
\end{equation}
\begin{equation}
{{\bf{\Omega }}_0} = \frac{1}{{8\pi {\eta _\infty }{a^3}}}\int_{{S_p}} {{{\bf{u}}^s} \cdot \left( {{\bf{n}} \cdot {{{\bf{\hat T}}}_\Omega }} \right)dS}  = {\bm{0}}.
\end{equation}

At first order, we have
\begin{equation}
{{\bf{U}}_1} = \frac{1}{{6\pi {\eta _\infty }a}}\int_{{S_p}} {{\bf{n}} \cdot {{\bm{\tau }}_{NN,1}}\,dS}  + \frac{1}{{6\pi {\eta _\infty }a}}\int {\left( {\nabla  \cdot {{\bm{\tau }}_{NN,1}}} \right) \cdot {{{\bf{\hat G}}}_U}\,dV} ,
\label{eqn:U1}
\end{equation}
\begin{equation}
{{\bf{\Omega }}_1} = \frac{1}{{8\pi {\eta _\infty }{a^3}}}\int_{{S_p}} {{\bf{r}} \times \left( {{\bf{n}} \cdot {{\bm{\tau }}_{NN,1}}} \right)\,dS}  + \frac{1}{{8\pi {\eta _\infty }{a^3}}}\int {\left( {\nabla  \cdot {{\bm{\tau }}_{NN,1}}} \right) \cdot {{{\bf{\hat G}}}_\Omega }\,dV.}
\label{eqn:Omega1}
\end{equation}
Here $\epsilon {{\bm{\tau }}_{NN,1}} = \left( {\eta \left( {\bf{x}} \right) - {\eta _\infty }} \right){{\bm{\dot \gamma }}_0} = \left( {\epsilon {\eta _1} + \eta '} \right){{\bm{\dot \gamma }}_0}$, where ${{\bm{\dot \gamma }}_0} = \nabla {{\bf{u}}_0} + {\left( {\nabla {{\bf{u}}_0}} \right)^T}$ is the rate of strain tensor associated with the leading order flow ${{\bf{u}}_0}$. We see that the variation of viscosity from the constant reference viscosity $\eta_{\infty}$ changes the traction locally on the surface of the active particle but also modifies the flow from its value in the homogeneous fluid ${{\bf{u}}_0}$. Both these changes alter the swimming velocity from that in the homogeneous fluid ${{\bf{U}}_0}$, ${{\bf{\Omega }}_0}$ represented by an extra velocity ${{\bf{U}}_1}$, ${{\bf{\Omega }}_1}$ (see Eqs.~\eqref{eqn:U1}, \eqref{eqn:Omega1}). The effect of locally modifying the viscosity from that of a constant viscosity $\eta_{\infty}$ to $\epsilon {\eta _1} + \eta '$ alone (without the effect of the associated change in the flow from ${{\bf{u}}_0}$) is referred to as the local effect in line with the literature on the motion of particles in shear-thinning fluids and is given by the surface integral terms in Eqs.~\eqref{eqn:U1}, \eqref{eqn:Omega1}. The complementary effect due to modifying the flow from the homogeneous fluid value ${{\bf{u}}_0}$ (without the associated change in the viscosity on the surface of the particle) is referred to as the non-local effect and this is given by the volume integral terms in Eqs.~\eqref{eqn:U1}, \eqref{eqn:Omega1}.

If the properties of the particle are same as those of the fluid, the particle does not produce any disturbance $\left( \eta' = 0 \right)$ and it propels with the first order velocities \citep{Datt2019}
\begin{equation}
{{\bf{U}}_1} =  - \frac{{a{B_2}}}{5}\left( {{\bf{I}} - 3{\bf{pp}}} \right) \cdot \nabla \left( {\frac{{{\eta _1}}}{{{\eta _\infty}}}} \right),
\label{eqn:U1-ambient}
\end{equation}
\begin{equation}
{{\bf{\Omega }}_1} =  - \frac{1}{2}{{\bf{U}}_N} \times \nabla \left( {\frac{{{\eta _1}}}{{{\eta _\infty}}}} \right).
\label{eqn:Omega1-ambient}
\end{equation}
The effect of the ambient viscosity is to turn the particle to align it against the viscosity gradient (negative viscotaxis). In this steady-state orientation, relative to their speed in the homogeneous fluid, the pushers speed up, the pullers slow down and the neutral swimmers do not experience any speed change. We next demonstrate the significance of the disturbance viscosity.

\section{\label{sec:Dist_Visc}Effect of the disturbance viscosity}
Recall the viscosity field satisfies the Laplace equation and $\eta  = {\eta _0} + \eta '$. As the ambient viscosity $\eta_0$ is linear in position, the disturbance viscosity $\eta'$ must also satisfy a Laplace equation
\begin{equation}
\nabla^2 \eta' = 0,
\label{eqn:Lap-dist-visc}
\end{equation}
and also decays to zero far from the particle
\begin{equation}
\eta ' = 0\,\,\,{\rm{as}}\,\,r \to \infty.
\label{eqn:dist-visc-far}
\end{equation}
The general solution of \eqref{eqn:Lap-dist-visc}, satisfying \eqref{eqn:dist-visc-far}, is given by 
\begin{equation}
\eta ' = \sum\limits_{k = 0}^\infty  {\sum\limits_{m = 0}^k {{r^{ - k - 1}}\left( {{A_{k,m}}\cos m\theta  + {B_{k,m}}\sin m\theta } \right)P_k^m\left( {\cos \phi } \right)} ,} 
\end{equation}
where $A_{k,m}$, $B_{k,m}$ are the constant coefficients, $\phi$, $\theta$ are the polar and azimuthal angles in the spherical coordinate system located at the center of the particle while $P_k^m$ is the associated Legendre polynomial of degree $k$ and order $m$. Here, the terms proportional to $1/r$, $1/r^2$, respectively, are the source and source-dipole, essentially the Green's functions of the Laplace equation and its derivative. The exact structure of the disturbance viscosity and its influence on the swimming velocities depend on the boundary condition on the particle surface. This surface can be impermeable (resp. insulating) to the nutrient or the salt concentration (resp. to the temperature) that is responsible for the viscosity variations. In this case, a no-flux condition for the viscosity holds on the particle surface, ${\bf{n}} \cdot \nabla \eta  = 0$ at $r = a$. Alternatively, the particle surface can be at a constant temperature which translates to a constant viscosity condition on the particle surface,  $\eta  = {\eta _p} = {\rm{constant}}$ at $r = a$. We hereby analyze these two cases separately.

\subsection{No-flux}
\begin{figure}[t!]
\subfloat{\includegraphics[scale = 0.5]{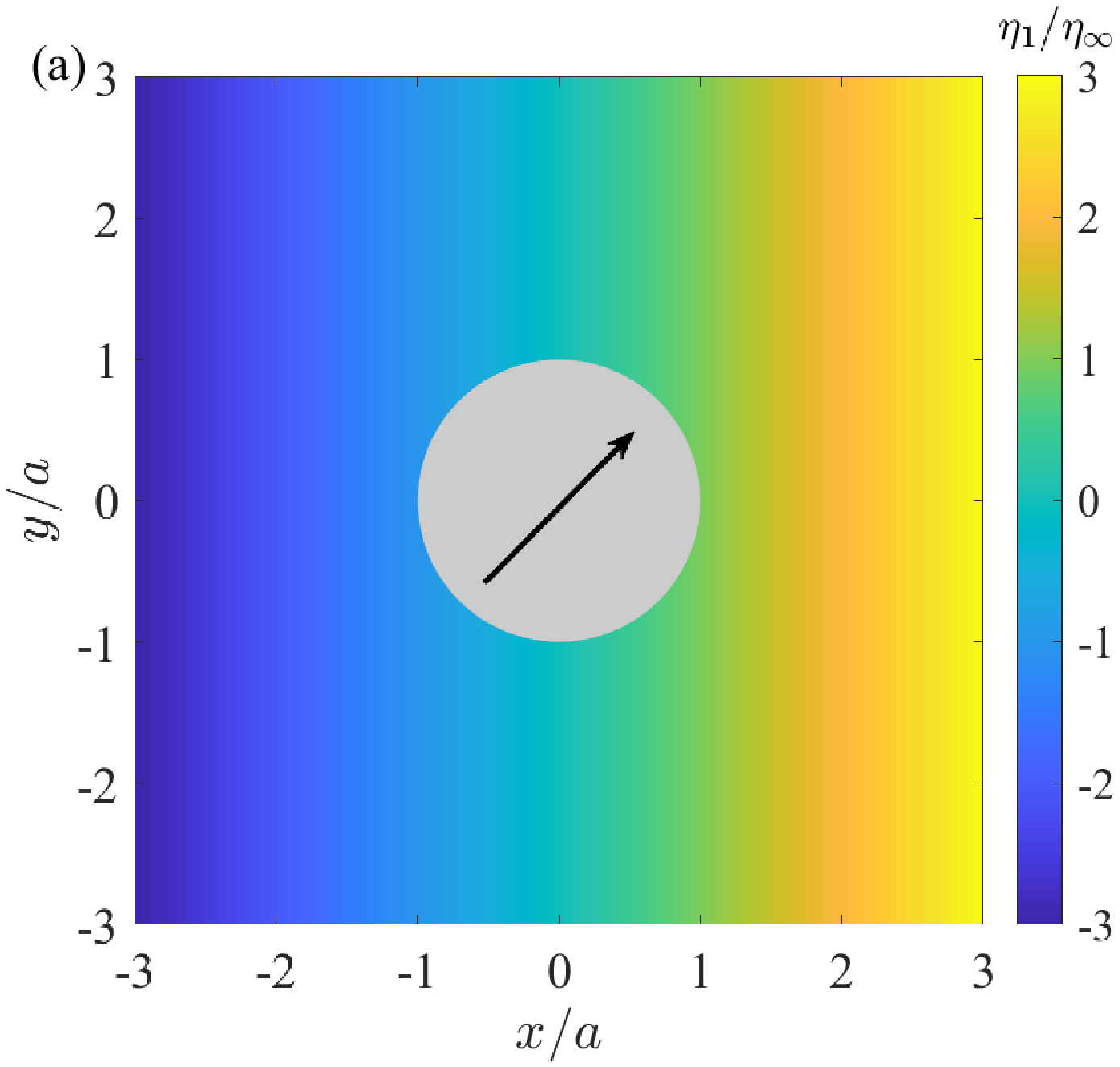}\label{fig:Amb_Visc}}
\hfill
\subfloat{\includegraphics[scale = 0.5]{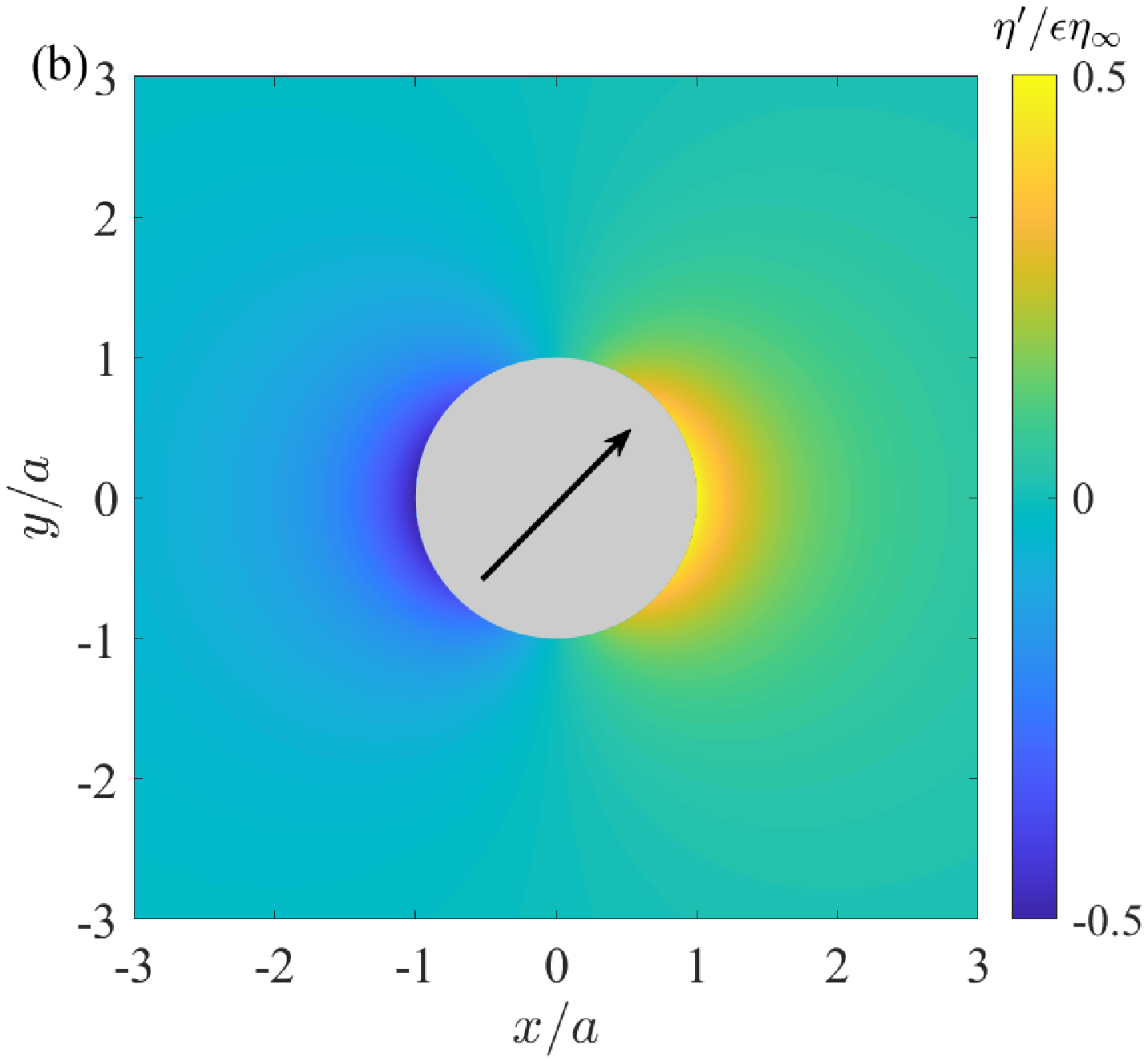}\label{fig:Dist_Visc_NF}}
\caption{\label{fig:Visc1}(Color online) (a) The ambient viscosity relative to the reference viscosity and (b) the disturbance viscosity due to the no-flux condition at the instant when the particle's center is at the origin in the lab frame i.e., ${{\bf{x}}_c} = {\bm{0}}$. These viscosity distributions do not depend on the orientation of the particle.}
\end{figure}

The no-flux condition can be simplified to
\begin{equation}
{\left. {\frac{{\partial \eta '}}{{\partial r}}} \right|_{r = a}} =  - \frac{{\epsilon {\eta _\infty }}}{a}\sin \phi \cos \theta.
\end{equation}
The disturbance viscosity that satisfies this constraint is given by
\begin{equation}
\eta ' = \frac{{\epsilon {a^2}{\eta _\infty }}}{{2{r^3}}}\left( {x - {x_c}} \right) = \frac{{\epsilon {a^3}}}{{2{r^3}}}{\eta _1 \left( x - x_c \right)}.
\label{eqn:Dist_Visc_NF}
\end{equation}
The no-flux condition prevents the occurrence of the source term and the solution is a source-dipole. This disturbance viscosity has the same angular dependence as that of the ambient viscosity, however, unlike the ambient viscosity, the disturbance viscosity decays with position (see Fig.~\ref{fig:Visc1}). This leads to a swimming velocity due to the disturbance viscosity of a similar form as that due to the ambient viscosity but with a reduced magnitude. The change in the velocity of the particle due to the disturbance viscosity alone (denoted by a prime) is given by 
\begin{equation}
{{\bf{U}}'_1} =  - \frac{{a{B_2}}}{{60}}\left( {{\bf{I}} - 3{\bf{pp}}} \right) \cdot \nabla \left( {\frac{{{\eta _1}}}{{{\eta _\infty }}}} \right),
\label{eqn:U1-no-flux}
\end{equation}
\begin{equation}
{{\bf{\Omega }}'_1} =  - \frac{1}{8}{{\bf{U}}_N} \times \nabla \left( {\frac{{{\eta _1}}}{{{\eta _\infty }}}} \right).
\label{eqn:Omega1-no-flux}
\end{equation}
Combining the contribution of the ambient and the disturbance viscosities, the change in the swimming velocity of a particle with no-flux condition is
\begin{gather}
{{\bf{U}}_1} =  - \frac{{13}}{{60}}a{B_2}\left( {{\bf{I}} - 3{\bf{pp}}} \right) \cdot \nabla \left( {\frac{{{\eta _1}}}{{{\eta _\infty }}}} \right),\\
{{\bf{\Omega }}_1} =  - \frac{5}{8}{{\bf{U}}_N} \times \nabla \left( {\frac{{{\eta _1}}}{{{\eta _\infty }}}} \right).
\end{gather}
We see that the disturbance viscosity does not change the physics predicted by considering only the ambient viscosity, it merely increases the rate at which the particle rotates to align opposite to the viscosity gradients and it also increases the speed change experienced by the pushers and pullers in the steady-state orientation (pushers swim faster and pullers swim slower while neutral swimmers propel at the same speed compared to that in the homogeneous fluid).

\subsection{Constant Viscosity}
The constant viscosity condition simplifies to
\begin{equation}
{\left. {\eta '} \right|_{r = a}} = {\eta _c} - \epsilon {\eta _{\infty}}\sin \phi \cos \theta,
\end{equation}
where ${\eta _c} = {\eta _p} - {\eta _\infty } - \epsilon {\eta _\infty }{x_c}/a$. This constraint has two parts, the first part is constant, $\eta_c$, while the second varies spatially over the particle surface. The constant is enforced through a source term while the spatial variation is satisfied through a source-dipole term. Putting these two together, the disturbance viscosity is
\begin{figure}[t!]
\subfloat{\includegraphics[scale = 0.5]{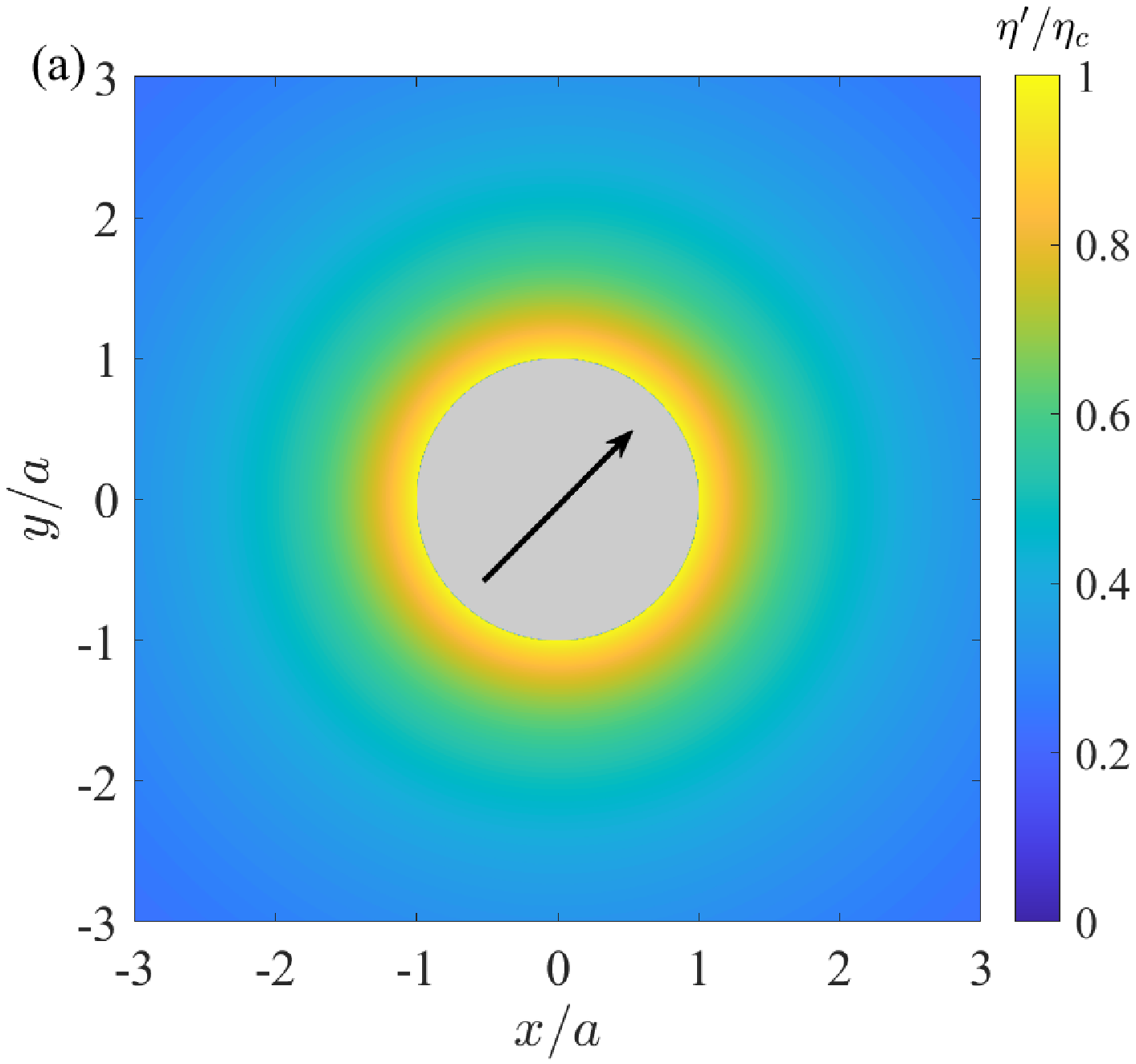}\label{fig:Dist_Visc_CV_S}}
\hfill
\subfloat{\includegraphics[scale = 0.5]{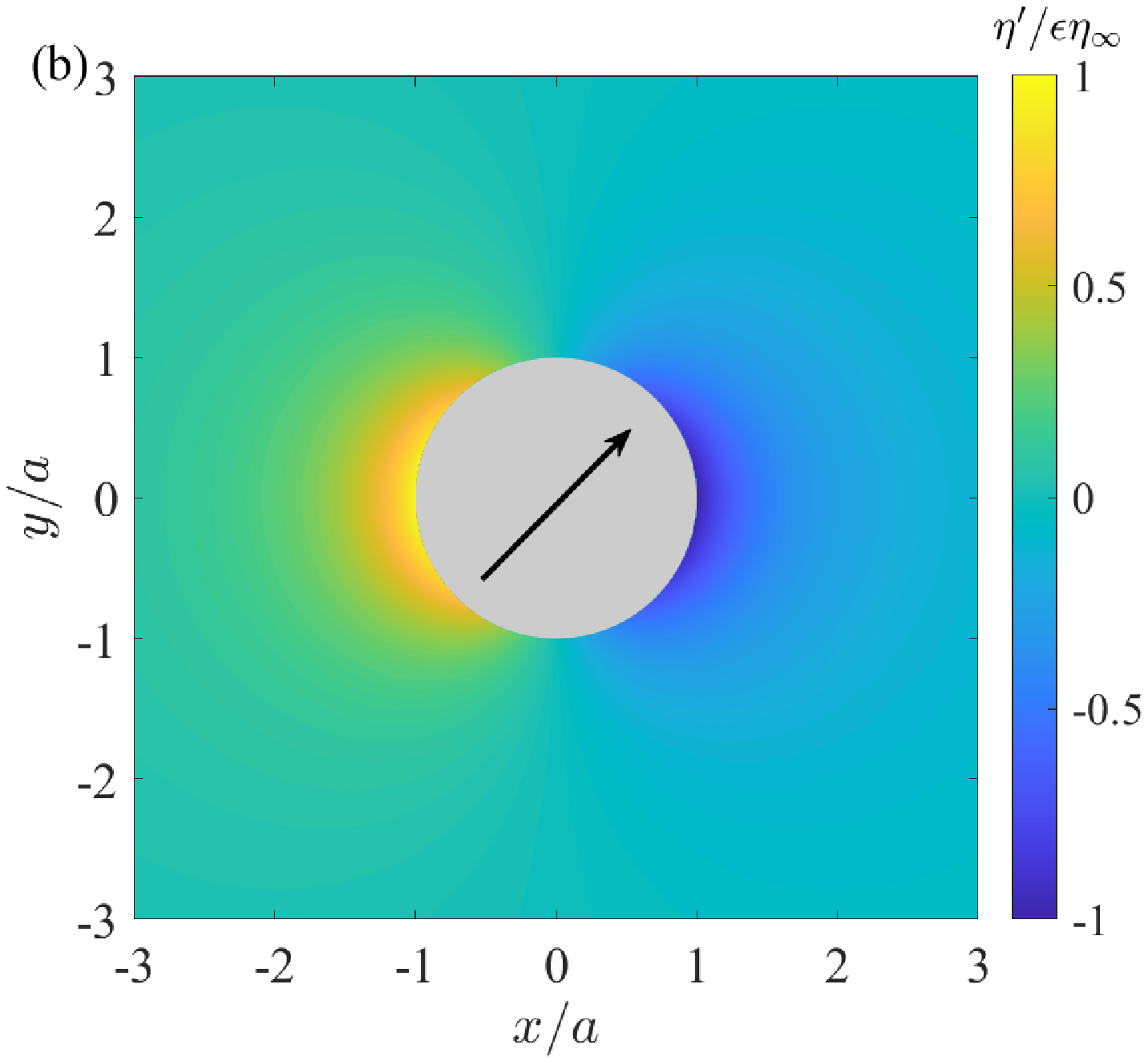}\label{fig:Dist_Visc_CV_SD}}
\caption{\label{fig:Visc2}(Color online) (a) The source and (b) the source-dipole part of the disturbance viscosity due to a constant viscosity condition at the instant when the particle center is at the origin in the lab frame i.e., ${{\bf{x}}_c} = {\bm{0}}$. These viscosity distributions do not depend on the orientation of the particle.}
\end{figure}
\begin{equation}
\eta ' = \frac{{{\eta _c}a}}{r} - \frac{{\epsilon {a^2}{\eta _\infty }}}{{{r^3}}}{\left( x - x_c \right)} = \frac{{{\eta _c}a}}{r} - \frac{{\epsilon {a^3}}}{{{r^3}}}{\eta _1 \left( x - x_c \right)}.
\label{eqn:Dist_Visc_CV}
\end{equation}
We see that the source dipole here is a factor $\left( { =  - 2} \right)$ times the source dipole due to the no-flux condition in \eqref{eqn:Dist_Visc_NF}, hence the swimming velocity due to the source dipole here should be the same factor times the swimming velocity due to the no-flux condition. The source term is spherically symmetric (see Fig.~\ref{fig:Dist_Visc_CV_S}), so it has no preferred direction. Hence the source term will induce only a translational velocity along $\bf{p}$ and no angular velocity. Overall, the change in the velocity of the particle due to the disturbance viscosity alone (with the constant viscosity boundary condition) is given by 
\begin{equation}
{{\bf{U}}'_1} = \frac{1}{{12}}\frac{{{\eta _{c}}}}{{\epsilon {\eta _\infty }}}{{\bf{U}}_N} + \frac{{a{B_2}}}{{30}}\left( {{\bf{I}} - 3{\bf{pp}}} \right) \cdot \nabla \left( {\frac{{{\eta _1}}}{{{\eta _\infty }}}} \right),
\label{eqn:U1-constant-eta}
\end{equation}
\begin{equation}
{{\bf{\Omega }}'_1} = \frac{1}{4}{{\bf{U}}_N} \times \nabla \left( {\frac{{{\eta _1}}}{{{\eta _\infty }}}} \right).
\label{eqn:Omega1-constant-eta}
\end{equation}
Again, summing up the contributions of the ambient and the disturbance viscosities, the change in the velocity of a particle with the constant viscosity boundary condition is
\begin{equation}
{{\bf{U}}_1} = \frac{1}{{12}}\frac{{{\eta _c}}}{{\epsilon {\eta _\infty }}}{{\bf{U}}_N} - \frac{{a{B_2}}}{6}\left( {{\bf{I}} - 3{\bf{pp}}} \right) \cdot \nabla \left( {\frac{{{\eta _1}}}{{{\eta _\infty }}}} \right),
\end{equation}
\begin{equation}
{{\bf{\Omega }}_1} =  - \frac{1}{4}{{\bf{U}}_N} \times \nabla \left( {\frac{{{\eta _1}}}{{{\eta _\infty }}}} \right).
\end{equation}
The disturbance viscosity does not change the reorientation process predicted by considering only the ambient viscosity, it only decreases the rate at which the particle turns to align itself against the viscosity gradients. The disturbance viscosity, however, modifies the speed changes experienced by the swimmer in the steady-state orientation in comparison to those predicted by considering the ambient viscosity alone. A neutral swimmer propels faster than that in the homogeneous fluid if it is cooler than the ambient fluid $\left( \eta_c > 0 \right)$. A pusher or puller swims faster or slower than that in the homogeneous fluid depending on the magnitude of $\alpha = B_2/B_1$ and how hot or cold the particle is relative to the ambient fluid.

%\textcolor{red}{The propulsion in the ambient or the disturbance viscosity can be understood by locally modifying the viscosity without changing the flow (the local effect) as explained next.}

\section{\label{sec:LocalNL}Local vs non-local effects}
Instead of analyzing local and non-local effects simultaneously, as we did in the previous section, in this section we look at each term separately and then quantify their relative importance. We do so because (as we show below) local effects are typically easy to predict, but alone may or may not qualitatively capture the dynamics.

Analyzing the results in Tables \ref{tab:table1}, \ref{tab:table2} and \ref{tab:table3} we find by comparing the local effects with the total response that the results are qualitatively similar for a particle imposing the no-flux condition to the viscosity field. However, for an active particle with a constant-viscosity boundary condition it is the non-local effects that are dominant. In fact, the local effects for such particle are identically zero. This is because the viscosity field $\epsilon {\eta _1} + \eta '$ near such particle is a constant $\left( { = {\eta _c}} \right)$ and this reduces the local effects to the force or torque acting on the particle in homogeneous fluid which is identically zero. In the sections below we unpack and parse these results in order to develop a more mechanistic understanding of the results in these tables.

\begin{table*}[t!]
\caption{\label{tab:table1}The first order local, non-local, and total translational velocity of a swimmer due to the $B_1$ squirming mode is given by $c\frac{{{\eta _c}}}{{\epsilon {\eta _\infty }}}{{\bf{U}}_N}$. This table provides the values of $c$ for various viscosity fields.}
\begin{ruledtabular}
\begin{tabular}{ccccc}
&Local&Non-local&Total\\ \hline
%& $\epsilon {\eta _1}$ & 0 & 0 & 0\\ \hline
\multirow{1}{*}{No-flux} 
%& $\eta'$ & 0 & 0 & 0\\
& 0 & 0 & 0\\ \hline
\multirow{1}{*}{Constant viscosity} 
%& $\eta'$ & 0 & $1/12$ & $1/12$\\
& 0 & $1/12$ & $1/12$\\
\end{tabular}
\end{ruledtabular}
\end{table*}

\begin{table*}[t!]
\caption{\label{tab:table2}The first order local, non-local, and total angular velocity of a swimmer with only $B_1$ squirming mode is given by $ - c{{\bf{U}}_N} \times \nabla \left( {\frac{{{\eta _1}}}{{{\eta _\infty }}}} \right)$. This table provides the values of the constant coefficient $c$ for various viscosity fields.}
\begin{ruledtabular}
\begin{tabular}{ccccc}
&Local&Non-local&Total\\ \hline
%& $\epsilon {\eta _1}$ & $1/2$ & 0 & $1/2$\\ \hline
\multirow{1}{*}{No-flux} 
%& $\eta'$ & $1/4$ & $-1/8$ & $1/8$\\
& $3/4$ & $-1/8$ & $5/8$\\ \hline
\multirow{1}{*}{Constant viscosity} 
%& $\eta'$ & $-1/2$ & $1/4$ & $-1/4$\\
& $0$ & $1/4$ & $1/4$\\
\end{tabular}
\end{ruledtabular}
\end{table*}

\begin{table*}[t!]
\caption{\label{tab:table3} The first order local, non-local, and total translational velocity of a swimmer with $B_n$ $\left( {n \ge 2} \right)$ squirming modes is given by $- ca{B_2}\left( {{\bf{I}} - 3{\bf{pp}}} \right) \cdot \nabla \left( {\frac{{{\eta _1}}}{{{\eta _\infty }}}} \right)$. This table provides the values of the constant viscosity coefficient $c$ for various viscosity fields.}
\begin{ruledtabular}
\begin{tabular}{ccccc}
&Local&Non-local&Total\\ \hline
%& $\epsilon {\eta _1}$ & $2/45$ & $7/45$ & $1/5$\\ \hline
\multirow{1}{*}{No-flux} 
%& $\eta'$ & $1/45$ & $-1/180$ & $1/60$\\
& $1/15$ & $3/20$ & $13/60$\\ \hline
\multirow{1}{*}{Constant viscosity}
%& $\eta'$ & $-2/45$ & $1/90$ & $-1/30$\\
& $0$ & $1/6$ & $1/6$\\
\end{tabular}
\end{ruledtabular}
\end{table*}

\subsection{Reorientation}
For simplicity, we consider a particle with only $B_1$ squirming mode and oriented orthogonal to the viscosity gradients along the ${\bf{e}}_y$ direction.

\subsubsection{Local effects}
The torque due to local changes in the viscosity is given by 
\begin{align}
{{\bf{L}}_{NN}^l}= \frac{{2{B_1}}}{a} \int_{{S_p}}\left( {\eta \left( {\bf{x}} \right) - {\eta _\infty }} \right){\bf{r}}\times{\bf{p}} dS.
\label{eq:local1}
\end{align}
From this we see that the viscosity at the surface of the particle simply acts as a weight on the moment arm, and that an asymmetry in the viscosity at the surface will lead to a reorientation of the particle towards the lower viscosity. The viscosity at the surface has a contribution from both the ambient and the disturbance $ {\eta \left( {\bf{x}} \right) - {\eta _\infty }} = {\epsilon {\eta _1} + \eta '}$. We know from \eqref{eqn:Dist_Visc_CV} that on the surface of the particle $\eta'=-\epsilon\eta_1+const.$ in order to satisfy a constant viscosity boundary condition, and it is clear from \eqref{eq:local1} that a constant viscosity on the surface of the particle clearly produces no net torque. In contrast, \eqref{eqn:Dist_Visc_NF} shows that in order to satisfy the no-flux boundary condition, $\eta'=\frac{1}{2}\epsilon\eta_1+const.$ meaning the asymmetry of the ambient viscosity at the surface of the particle, and thus the torque, is enhanced by a 3/2 factor (see Fig.~\ref{fig:Dist_Visc_NF}). An application of the divergence theorem to \eqref{eq:local1} allows one to simply resolve the torque as
\begin{align}
{\bf{L}}_{NN}^l &= -\frac{9}{2}{\bf{U}}_N\times \int_{V_p} \epsilon\nabla\eta_1 dV = -6\pi a^3{\bf{U}}_N\times\epsilon\nabla {\eta _1},
\end{align}
and multiplication by the mobility leads to the result in Table \ref{tab:table2}.

  \subsubsection{Non-local effects}
Recall the non-local effect is given by $\int {\left( {\nabla  \cdot {{\bm{\tau }}_{NN,1}}} \right) \cdot {{{\bf{\hat G}}}_\Omega }} \,dV$. We use the expression for ${{\bf{\hat G}}_\Omega }$ to simplify this integral to 
\begin{equation}
\int {\left( {\frac{{{a^3}}}{{{r^3}}}} \right){\bf{r}} \times \nabla  \cdot \left[ {\left( {\epsilon {\eta _1} + \eta '} \right){{{\bm{\dot \gamma }}}_0}} \right]dV}.
\end{equation}
As the strain-rate tensor due to $B_1$ mode is divergence-free, the non-local effect finally simplifies to
\begin{equation}
\int {\left( {\frac{{{a^3}}}{{{r^3}}}} \right){\bf{r}} \times \nabla \left( {\epsilon {\eta _1} + \eta '} \right) \cdot {{{\bm{\dot \gamma }}}_0}dV}.
\end{equation}

We examine the non-local effect by analyzing the kernel of the integral that represents the non-local effects, $\nabla \left( {\epsilon {\eta _1} + \eta '} \right) \cdot {{\bm{\dot \gamma }}_0}$. The contribution of ambient viscosity to this kernel that is not along the radial direction is
\begin{equation}
\frac{{\epsilon {\eta _\infty }}}{a}\frac{{2{a^3}}}{{{r^5}}}{B_1}\left( {x - {x_c}} \right){{\bf{e}}_y} + \frac{{\epsilon {\eta _\infty }}}{a}\frac{{2{a^3}}}{{{r^5}}}{B_1}\left( {y - {y_c}} \right){{\bf{e}}_x}.
\end{equation}
Half of this kernel (the first term) has left-right mirror symmetry while the other half has front-back mirror symmetry. Each half of this kernel induces an angular velocity of equal magnitude but opposite direction making the net non-local effect due to the ambient viscosity to be zero. The source part of the disturbance viscosity as well as its gradient do not have any directional preference or any asymmetries. Hence, the non-local effect due to the source is zero. The source dipole part of the disturbance viscosity is proportional to $x$ and it decays with the position. So, the gradient of the source dipole has two parts. One is along the gradient of the ambient viscosity and hence, like the ambient viscosity, this part induces zero angular velocity. The other part is in the radially inward direction due to the decay of source-dipole with position and also it is proportional to $x$. The latter characteristic introduces a left-right mirror symmetry to the kernel that consequently induces a non-zero angular velocity. As the disturbance viscosity always has a source-dipole component, its non-local effect is non-zero regardless of the boundary condition on the particle surface. See the third column in Table \ref{tab:table2}.

\subsection{Speed changes in the steady-state orientation}
The steady-state orientation is opposite to the ambient viscosity gradients. For the ambient viscosity, ${\eta _0} = {\eta _\infty } + \epsilon {\eta _\infty }x/a$, this orientation is along $ - {{\bf{e}}_x}$. In this configuration, the flow and the viscosity fields are axisymmetric, the axis of symmetry is along the swimmer's orientation. To understand the speed changes in the steady-state orientation, we consider a swimmer with first two squirming modes as the swimming velocity in viscosity gradients anyways depends on only these two squirming modes.
\subsubsection{Local effects}
The force due to local variation in the viscosity is given by
\begin{equation}
{\bf{F}}_{NN}^l = \int_{{S_p}} {{\bf{n}} \cdot {{\bm{\tau }}_{NN}}dS}  = \int_{{S_p}} {\left( {\epsilon {\eta _1} + \eta '} \right){\bf{n}} \cdot {{{\bm{\dot \gamma }}}_0}\,dS} .
\end{equation}
For a particle with constant viscosity condition, the viscosity $\epsilon {\eta _1} + \eta '$ is constant $\left( { = {\eta _c}} \right)$ near the particle and this reduces the local force to the force due to viscous traction in homogeneous Newtonian fluid which is identically zero. In homogeneous fluids, the force due to pressure and viscous traction are both zero so as to enforce the force-free condition on the particle. On the other hand, near a particle with a no-flux boundary condition, the viscosity $\epsilon {\eta _1} + \eta '$  is $\frac{3}{2}\epsilon {\eta _1} + const.$ Here, the ambient viscosity $\eta_1$ is front-back mirror symmetric as ${\eta _1} \propto x$. The flow and hence the shear-rate ${\dot \gamma _{0,rx}}$ due to a particle with only $B_1$ mode in the homogeneous fluid are front-back symmetric. This symmetry together with the mirror-symmetry in the viscosity induce a mirror-symmetry in the traction ${\bf{n}} \cdot {{\bm{\tau }}_{NN,1}}$, which means that for every traction in front of the particle, there is an equal in magnitude but oppositely directed traction on the back of the particle making the net local force to vanish. The flow and the shear-rate ${\dot \gamma _{0,rx}}$ generated by a particle with only $B_2$ mode in the homogeneous fluid are front-back mirror symmetric. This mirror-symmetry along with a similar symmetry in the viscosity makes the traction front-back symmetric, meaning that for every traction in front of the particle, there is an equal traction behind the particle resulting in a finite local force. See the second column in Tables \ref{tab:table1}, \ref{tab:table3}.

\subsubsection{Non-local effects}

We examine the non-local effects by analyzing the kernel of the integral that represents the non-local effects, $\nabla  \cdot {{\bm{\tau }}_{NN,1}} \cdot {{\bf{\hat G}}_U}$ or $\nabla  \cdot {{\bm{\tau }}_{NN,1}}$. Through a linear decomposition of the first order problem, we can show that the term $\nabla  \cdot {{\bm{\tau }}_{NN,1}}$ modifies the flow from that in the homogeneous fluid and therefore, is responsible for the non-local effect \citep{Pietrzyk2019}. If the kernel and hence, the modified flow are front-back mirror symmetric, then the non-local effect being the integral of the kernel vanishes.

A particle with only $B_1$ mode generates a front-back symmetric flow in the homogeneous fluid. Then the ambient viscosity and the source-dipole part of the disturbance viscosity introduce the mirror-symmetry to the kernel diminishing the non-local effect but the source part of the disturbance viscosity does not bring in such symmetry inducing a finite non-local effect. As a particle imposing a constant-viscosity condition generates a source disturbance, the non-local effect is non-zero for such particle but it is zero for a particle with no-flux condition. See the third column in Table \ref{tab:table1}.

In the case of a particle with only $B_2$ mode, the flow in the homogeneous fluid itself is front-back mirror symmetric. The additional mirror symmetry of the ambient viscosity or the source-dipole part of the disturbance viscosity prevents such symmetry in the kernel generating a finite non-local effect. But the kernel due to a source still has this mirror-symmetry yielding a zero non-local effect. These non-local effects due to the ambient or the disturbance viscosity manifest as a finite non-local effects on a particle with any boundary condition. See the third column in Table \ref{tab:table3}.

\section{\label{sec:Conc}Conclusions}
Spatial variations of viscosity in fluids may arise due to changes in temperature, salt or nutrient concentration. Insertion of a particle into the fluid tends to disturb the background viscosity field due to boundary conditions imposed by the particle on the viscosity field. In general we find that active squirmer-type particles reorient down the viscosity gradient (negative viscotaxis) and they propel in the steady-state orientation with a speed different from that in the homogeneous fluid. The specific boundary conditions on the viscosity field at the surface of the particle do not qualitatively affect the reorientation process, only changing the rate at which the steady-state orientation is achieved. On the contrary, the speed changes experienced by the swimmer while propelling along the steady-state orientation are significantly dependant on the boundary conditions on the particle. For active particles that impose a no-flux condition, pushers speed up, the pullers slow down and the neutral swimmers do not change their speed. In contrast the speed changes experienced by a particle that imposes a constant-viscosity boundary condition are more complex. Even the neutral swimmer speeds up if it is cold relative to the ambient fluid while the speed changes of a pusher or a puller depend on how strong the pusher or puller is (the ratio of the squirming modes, $B_2/B_1$) and how hot or cold the particle is relative to the ambient fluid. This suggests the additional degrees of freedom, the particle temperature and the squirming modes, that one can explore in controlling the active matter.

We also quantified the relative importance of the local and non-local effects, where the local effects are a consequence of locally modifying the viscosity but not the flow while the non-local effects are due to changing the flow from that in the homogeneous fluid without altering the viscosity. We found that the local effects are sufficient to determine the motion of a particle imposing a no-flux boundary condition but it is the non-local effects that dominate the (change in) dynamics of a particle imposing a constant-viscosity boundary condition. Hence, one should exercise caution when using a simple local effects based theory to study the motion of active particles in viscosity gradients.

\bibliography{references}
\end{document}